\documentclass[aps,prb,twocolumn,a4paper]{revtex4}
\usepackage{amsmath}
\usepackage{amssymb}
\usepackage{graphicx}
\usepackage{epsfig}

\begin{document}

\title{Density matrix renormalization group for bosonic quantum Hall effect}

\author{D.~L. Kovrizhin}
\affiliation{Theoretical Physics, Oxford University, 1 Keble Road, OX1 3NP, Oxford, United Kingdom}

\begin{abstract}
We developed a density matrix renormalization-group technique to study quantum Hall fractions of fast rotating bosons. In this paper we present a discussion of the method together with the results which we obtain in three distinct cases of the  narrow channel, cylinder and spherical geometries. In the narrow channel case, which is relevant to anisotropic confining traps in the limit of extremely fast rotation, we find a series of zero-temperature phase transitions in the strongly interacting regime as a function of the interaction strength between bosons. We compute energies and density profiles for different filling fractions on a cylinder and compare the convergence rates of the method in the cylinder and a sphere geometries.
\end{abstract}

\maketitle

\section{Introduction}

Density matrix renormalization-group (DMRG) methods play an important role in numerical studies of one-dimensional quantum
lattice models allowing to access their low-energy properties with a very high precision.\cite{review} Recently, these powerful techniques have also been successfully applied to nonequilibrium problems in strongly-interacting quantum systems  such as dynamics of quantum quenches \cite{dynamics} and tunneling of electrons through quantum point contacts.\cite{fendley} A considerable theoretical effort has been put into development of similar methods for two-dimensional quantum lattice models, see for example Refs. \cite{vidal} and \cite{tensor} These new numerical techniques make use of the recent advances in understanding entanglement properties of many-particle quantum systems in finding the optimal basis for the representation of the ground state and low-lying excitations. This is usually implemented by using matrix-product states or tensor networks. \cite{tensor} DMRG methods have also been used in quantum chemistry and nuclear physics.\cite{papenbrock}

An interesting application of DMRG was developed by Shibata and Yoshioka\cite{shibata} and more recently by Feiguin \textit{et al.} in the studies of interacting electrons in the fractional quantum Hall effect (FQHE) regime. In the paper in Ref. \cite{feiguin_recent}, the authors calculated ground state energies, gaps and correlation functions of electrons at $\nu=1/3$ and $\nu=5/2$ filling fractions in the spherical geometry for systems with larger number of particles than it was possible to study with exact diagonalization. These calculations showed a potential of the method in finite-size numerical studies of quantum Hall systems. A standard approach in FQHE is exact diagonalization which has been very fruitful in identifying states of interacting electrons with different filling fractions and studying excitation spectra in these states. However, in some of the systems of current theoretical and experimental interest, exact diagonalization has almost reached its limits because of the exponential growth of Hilbert-space dimension with the system size. There are many examples where a new method would be beneficial, such as QHE systems with spin degrees of freedom and double-layer configurations, studies of higher filling fractions when several Landau levels are occupied as well as problems related to nonequilibrium dynamics of electrons in quantum Hall edge states which are still far from being well understood, see for example Ref.\cite{neder}

Recent advances in experiments with cold atomic gases allow to study systems of neutral particles, both fermions and bosons, under rotation, which is equivalent to having a coupling to a magnetic field in the rotating frame. These experiments showed formation of the triangular Abrikosov vortex lattice in rotating Bose-Einstein condensates and in systems of fermions with attractive interactions.\cite{ketterle}  A fast rotation regime was achieved experimentally,\cite{cornell} where disappearance of the vortex lattice was observed at very high rotation frequency. Alternative way of creating artificial magnetic fields based on the use of optical lattices was proposed in Ref. \cite{osterloh} and has been recently realized in experiment\cite{lin}. An interesting question about melting of the vortex lattice under increased rotation was investigated by Cooper \textit{et al.}\cite{cooper}. Quantum Hall fractions of rotating bosons have been extensively studied with exact diagonalization by Regnault \textit{et al.}\cite{regnault} Critical properties and the phase diagram of bosons in extremely elongated traps (narrow channel geometry) were discussed recently by Matveenko \textit{et al.}\cite{matveenko}

In this paper we report on the DMRG technique for rotating \textit{bosons} in the quantum Hall effect regime. This has been possible because of the special structure of the Hamiltonian projected onto the lowest Landau level which becomes essentially one-dimensional in the momentum space with long-range interactions generated by the projection. It is well-known that DMRG works better in systems with short-range interactions, however there are successful applications of the method to calculations for strips of finite width.\cite{ladders} Moreover in the case of quantum Hall effect in a cylinder geometry the matrix elements of interactions between different momentum eigenstates are exponentially suppressed at large transferred momenta. We discuss convergence of the method in spherical geometry and on a cylinder for different filling fractions.

The structure of the paper is the following. In Sect. II we introduce the Hamiltonian describing a system of
fast-rotating bosons. In Sect. III we explain our method on an example of the narrow-channel geometry. In Sect. IV we present the results of our calculations for systems in spherical and cylinder geometries.

\section{Bosons under extreme rotation}
Let us consider a system of neutral bosonic atoms, which are strongly confined in the $z$ direction by an external trapping potential with the frequency $\omega_z$ such that the bosons are in the ground state of the harmonic well. In this limit the system becomes essentially two-dimensional in the $(x,y)$ plane. The bosons are confined in this plane by an additional harmonic trapping potential $V({\bf r})$, with ${\bf r}=\{x,y\}$ which rotates around the $z$ axis with a frequency $\mathbf{\Omega}=\Omega \mathbf{\hat z}$. A single-particle Hamiltonian for this systems is equivalent to one describing a particle with an effective charge $q^{*}$ and a mass $m$ in an uniform magnetic field pointing along the $z$ axis. In the case of a symmetric harmonic trap with a frequency $\omega$ and $V(\mathbf{r})=m\omega^2\mathbf{r}^2/2$ this Hamiltonian reads
\begin{equation}
\hat{H}=\frac{1}{2 m}{(\hat{\mathbf p} - q^{*} {\bf{A}} /c)^2} + \frac{1}{2}m (\omega^2 -\Omega^2)  \mathbf{r}^2.
\label{Hsingle}
\end{equation}
Here $\mathbf{p}$ is a two-dimensional momentum operator in the $(x,y)$ plane, $\mathbf{A}=(m c/q^{*})[\mathbf{\Omega}\times\mathbf{r}]$ is a vector potential
which corresponds to an effective magnetic field $\mathbf{B}=[\mathbf{\nabla}\times\mathbf{A}]=2m\Omega c/q^{*}\hat{\mathbf{z}}$ and $c$ is a speed of light.
At the critical rotation frequency $\Omega=\omega$, the residual confinement vanishes and the dynamics is governed by the Hamiltonian of a charged spinless particle moving in an infinite two-dimensional plane in external magnetic field.
\begin{equation}
 \hat H_{p}=\frac{1}{2m}(\mathbf{\hat p}-q^{*}\mathbf A/c)^2.
\end{equation}

In the following we will study systems of bosonic atoms with short-range delta-function interactions
\begin{equation}
\hat{H}_{int}^{(3d)}(\mathbf r-\mathbf r')=g_{3d}\delta^{(3d)}(\mathbf r-\mathbf r'),\label{dint}
\end{equation}
with $g_{3d}=4\pi \hbar^2 a_s/m$, where $a_s$ is the three-dimensional scattering length.
If the harmonic oscillator length in the $z$ direction, $l_z=\sqrt{\hbar/m\omega_z}$, is much larger than the scattering length $|a_s|$ and the characteristic radius of interparticle interaction, one can find for the effective interaction constant \cite{gora1}
\begin{equation}
g_{2d}=\frac{2\sqrt{2\pi}\hbar^2a_s}{ml_z}.
\label{g}
\end{equation}
The many-particle Hamiltonian of the system in the quasi-two-dimensional geometry reads
\begin{equation}
 \hat{H}_{2d}=\sum_{a}\frac{1}{2m}(\hat{\mathbf{p}}_{a}-q^{*}\mathbf A_a/c)^2+\sum_{a<b}g_{2d}\delta^{(2d)}(\mathbf{r}_{a}-\mathbf{r}_b)
\label{ham_magn}
\end{equation}

When the chemical potential of the gas $\mu\sim g_{2d} n_p$ with the density $n_p$, is much smaller than the cyclotron gap $\hbar\omega_c=\hbar q^{*}B/mc=2\hbar\Omega$, the system can be effectively described by projecting it to the lowest Landau level. In this case the kinetic energy is quenched to zero and the dynamics is governed by interactions which leads to ground states depending on the boson filling fraction $\nu$, which is given by the ratio of the number
of particles $N_p$ to the number of flux quanta $N_v=A/2\pi l^2$, where $A$ is the area of the system and
$l=\sqrt{\hbar c/q^{*} B}=\sqrt{\hbar/2m\Omega}$ is a magnetic length. In the limit of large filling fractions the ground state is gapless and is given by an Abrikosov vortex lattice. At small filling fractions it is represented by strongly correlated states which are gapped. The Abrikosov lattice and the gapped states are connected at intermediate filling fractions of the order of $\nu\sim7$ by a phase transition with melting of the vortex lattice. \cite{cooper}

\section{Discussion of the DMRG method}

Density matrix renormalization group is a standard method, which was initially developed for one-dimensional quantum lattice models by White in Ref. \cite{white} It allows to calculate various quantum mechanical
observables such as energies, correlation functions, etc. for the ground and a few excited states of the systems
with short-range interactions with an extremely high precision. In the following we will extend the standard DMRG approach to study systems of rotating \textit{bosons} described by the projected to the lowest Landau level Hamiltonian (\ref{ham_magn}). DMRG for fermions in the QHE regime have been studied in Refs. \cite{feiguin,feiguin_recent}

The general idea of the method is to use the eigenvectors, corresponding to the highest eigenvalues of the density matrix,
which is calculated for a part of the system, as an optimal basis to represent a target state, usually the ground state.
There are two main versions of the algorithm i.e. the finite size and the infinite size
methods.\cite{white} To achieve best results one usually starts with the infinite size approach and correct it
afterwards by implementing finite size sweeps. In our case because of the bosonic statistics particle occupation numbers can be large, moreover one has to impose constraints on the total number of particles and the total momentum, which makes it difficult to use standard methods and we have to implement a different approach which is presented below. 

\subsection{Narrow-channel geometry}
We will discuss our method using an example of the narrow-channel geometry, which
was introduced\cite{sinha} and studied recently\cite{matveenko} in the context of rotating bosons
in strongly elongated traps in the mean-field regime.\cite{ho}
This situation arises when the confining potential in the $x$ direction is smaller than the trapping frequency
in the $y$-direction and the rotation frequency
$\Omega$ is equal to $\omega_x$. In this limit the system becomes infinitely elongated in the $x$ direction and
have remaining confinement in the $y$ direction. The single-particle Hamiltonian in the rotating frame in
the Landau gauge reads\cite{matveenko}
\begin{equation}
 \hat H_0=\frac{1}{2m}(\mathbf{\hat p}+2m\Omega y \mathbf{e}_x)^2+\frac{1}{2}m\omega_{-}^2 y^2,
 \label{narrow_ham}
\end{equation}
where $\omega_{-}^2=\omega_{y}^2-\Omega^2>0$.

We impose periodic boundary conditions on the system of finite length $L$ in the $x$ direction so that
the corresponding momentum is quantized as $k=2\pi n/L $ with $n\in\mathbb Z$ and assume that $\Omega\gg\omega_{-}$.
\cite{sinha}
The eigenstates of Hamiltonian (\ref{narrow_ham}) in the lowest Landau level are given
by plane-waves with momentum $k$ in the $x$ direction and have a gaussian profile in the $y$ direction which is centered at the positions shifted by the value of $kl^2$ with respect to the origin. These eigenfunctions read
\begin{equation}
 \varphi_k(x,y)=\frac{1}{\sqrt{L}}\frac{1}{(\pi l^2)^{1/4}} e^{ik x}e^{-\frac{1}{2l^2}(y-k l^2)^2}.
\label{phik}
\end{equation}
The set of states (\ref{phik}) with all possible momenta $k$ represents a basis on the lowest Landau level.

The many-particle Hamiltonian for bosons with delta-function interactions (\ref{dint}) in the narrow channel geometry is given by the equation
\begin{multline}
 \hat H_{nc} = \int dx dy\ \hat\Psi^{+}\left[(\mathbf{\hat p}+2m\Omega y \mathbf{e}_x)^2 + \frac{m\omega_{-}^2y^2}{2}\right]\hat\Psi\\ +\frac{g_{2d}}{2}\int dx dy\ \hat\Psi^{+}\hat \Psi^{+}\hat\Psi\hat\Psi.
\label{ham_xy}
\end{multline}
In order to project this Hamiltonian onto the lowest Landau level we write the operators $\hat\Psi(x,y)$ in the basis of eigenfunctions (\ref{phik})
\begin{equation}
 \hat\Psi(x,y)=\sum_{k=-\infty}^{\infty}\hat{a}_k \varphi_k(x,y).
\label{psi}
\end{equation}
where $\hat a_{k},\hat a_{k}^{+}$ are boson annihilation and creation operators with the commutation relations
$[\hat{a}_k,\hat{a}_{q}^{+}]=\delta_{kq}$. After substitution of Eq. (\ref{psi}) into Eq. (\ref{ham_xy}) and integration over $x,y$ we obtain
\begin{equation}
\hat{H}_{nc}=\sum_{k=-\infty}^{\infty}k^2 \hat{a}_{k}^{+}\hat{a}_{k}+
\frac{1}{2}\sum_{ijkl}V_{ijk}\hat a_i^{+}\hat a_j^{+}\hat a_k \hat a_l\delta_{i+j,k+l}.\label{ham_sec}
\end{equation}
Here the energy is given in units of $\hbar^2/2m^{*}l^2$ and $m_{*}=m(2\Omega/\omega_{-})^2$ is the effective mass.
The matrix elements of the interaction potential are given by the equation 
\begin{equation}
V_{ijk}=g e^{-\frac{l^2}{2}[(i-k)^2+(j-k)^2]},\label{matr_el}
\end{equation}
where $g=\sqrt{{2}/{\pi}}l{m^{*}g_{2d}}/{\hbar^2}L$ is the dimensionless coupling constant.

The properties of the ground states of the Hamiltonian (\ref{ham_sec}) depend on three parameters for the system of finite size, namely the coupling constant $g$, the length of the system $L$ and the total number of particles $N_p$. In the thermodynamic limit the length dependence disappears and we will only have two parameters, the interaction strength and the dimensionless linear particle density $n_p l=N_p l/L$. The limit $n_p l\ll1$ corresponds to a quasi-one-dimensional case and is similar to a Lieb-Liniger gas, the opposite limit $n_p l\gg1$ for small interactions (mean-field regime) was studied in.\cite{sinha,matveenko} Here we will consider another regime of strong interactions $g\gg1$ and large particle densities $n_p l\gg1$.

\subsection{DMRG algorithm for bosonic quantum Hall effect}

Our method is based on application of DMRG in the momentum space which was proposed by Xiang in the context of
the Fermi-Hubbard model\cite{xiang}. However, in our case the boson occupation numbers can take any value
from zero to the total number of particles in contrast to fermions and one has to modify the method of Ref.\cite{xiang}
In this subsection we will consider the system with Hamiltonian (\ref{ham_sec})
with total number of particles $N_p$ and length~$L$.

\begin{figure}[tb]
\epsfig{file=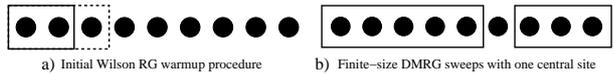,width=8cm}
\caption{ a) Warmup process using the standard Wilson renormalization procedure. A sweep is started from the leftmost site with one site added at each step on the right until we reach the central site.
b) Finite-size DMRG sweeps with a single central site after the initial Wilson RG procedure.}
\label{fig:0}
\end{figure}

We put the system on a lattice with a fixed number of sites $N_s=2N_{max}+1$ in momentum space with
$k=2\pi n/L$, where $n$ takes values $n=-N_{max}..N_{max}$ and $N_{max}$ is a cutoff. In general, the number of sites is infinite, but for the narrow-channel geometry there is always a physical cut-off due to the ``kinetic energy'' term in the Hamiltonian. The Hilbert space of the problem is represented by the states with the sites filled with bosons taking occupation numbers $n_k$ from $0$ to $N_p$. Hamiltonian (\ref{ham_sec}) conserves total number of particles $N_p=\sum{n_k}$ and total momentum $P=\sum{k n_k}$, which we implement as constraints on allowed basis states.

Step (I). We start with the Wilson renormalization group procedure from the leftmost site which has momentum $-N_{max}$. In the following we will use site numbers to denote the momenta. We take few sites $m$ with momenta from $-N_{max}$ to $-N_{max}+m-1$. In our calculations we used $m$ from one to three. We diagonalize the Hamiltonian in the basis generated on these sites in all possible subspaces $(n,p)$ of number of particles and momenta, where $n$ is an integer which runs from zero to a total number of particles and $p$ takes values from $-n\times N_{max}$ to $n\times(-N_{max}+m-1)$. We reduce the basis by omitting the states which can not contribute to the target state. For example if we are interested in the state with total momentum $P$, we should not take into account subspaces which have no partner $(n',p')$ on the sites which do not belong to a current block, i.e. if there is no set of $(n',p')$ such that $n+n'=N_p$ and $p+p'=P$. After diagonalizing the Hamiltonian in all the subspaces, we project it in every subspace together with the matrices corresponding to the matrix elements of the operators $\hat{a}_p$ between allowed bra and ket subspaces, onto the lowest energy eigenstates of the Hamiltonian. In the calculations we take up to four states in every subspace.  After projection we save the operator matrices on a computer hard disk.

Step (II). We add a single site to the block which is calculated in the previous step. When the first site is added it would be an initial block diagonalized in the Step I. This is done by generating new basis blocks $(n_{i},p_{i})$ from the states of the previous block and the states on the site with particle number and momentum conservation constraints as in the Step I, here $i$ is the current number of the Wilson RG step. We generate the Hamiltonian in the new combined block in all $(n_{i},p_{i})$ subspaces by taking a tensor product of the operators, represented by matrices, from the previous block and from the added site. We project the operators and the Hamiltonian to its lowest energy eigenstates as in Step I and save them to the hard disk.
In addition, every RG step we save the following set of operators
\begin{align}
\hat{A}_{0}(i)=\hat{a}_{i}, \hat{A}_{1}(i,j)=\hat{a}_{i}^{+}\hat{a}_{j},\\
\hat{A}_{2}(i,j)=\sum_{k}{\hat{a}_{j+k}^{+}\hat{a}_{i+k}e^{-k^2/2}},\\
\hat{A}_{3}(i)=\sum_{j,k}\hat{a}_{j}^{+}\hat{a}_{k}\hat{a}_{i+j-k}e^{-1/2[(i-k)^2+(j-k)^2]}.
\end{align}
This procedure suggested by Xiang in Ref. \cite{xiang} significantly reduces computation time.
We repeat Step II until we reach the central site. It is important to mention that we keep all
the matrix elements of the interactions in the Hamiltonian and allow for all possible boson occupation
numbers, at a given total number of particles, in the basis states. 

Step (III). In this step we start finite-size version of DMRG. First, we separate the Hamiltonian into two blocks and a single central site in the following way
\begin{equation}
\hat{H}=\hat{H}_{L}+\hat{H}_{L\bullet}+\hat{H}_{\bullet}+\hat{H}_{\bullet R}+\hat{H}_{R}+\hat{H}_{LR},
\end{equation}
where $L,R$ denote the left and the right block, the site is represented by the symbol $\bullet$.
The site Hamiltonian is diagonal and is given by the sum of the kinetic energy term and the part of the
interactions with operators acting on the site occupation numbers. The operators $\hat{H}_{L,R}$ represent
the contribution to the Hamiltonian from the left or the right block, the operators $\hat{H}_{LR}$ are composed from the
contributions connecting the left and the right blocks, $\hat{H}_{L\bullet}$ and  $\hat{H}_{\bullet R}$ connect the central site to the left or the right block correspondingly.

At every DMRG step we generate a new basis to obtain the set of states with the target quantum numbers $(N_p,P)$
from the possible states in the blocks which satisfy the following equations $N_p=N_L+N_{\bullet}+N_R$ and $P=P_L+P_{\bullet}+P_{R}$, where $N_{{L,R,\bullet}}, P_{{L,R,\bullet}}$ are the number of particles and momentum in a given subspace of the corresponding block or site.

The Hamiltonian for the right block is obtained by reflecting the Hamiltonian of the left block. We create the full Hamiltonian of the system by a tensor product of the matrices from the left block, the site and the right block in the basis calculated above. We use the eigenstates of this Hamiltonian to calculate the density matrix for the left or right part of the system by tracing out the coordinates of the remaining part in every subspace of quantum numbers. After diagonalizing the density matrix we generate a matrix of the eigenvectors with highest eigenvalues adding the extra eigenvectors if necessary to have at least one eigenvector in every subspace. We fix the maximum number of eigenvectors or select the ones which eigenvalues are greater than some small number $\varepsilon$. In our calculations we choose $\varepsilon$ in the range from $10^{-8}$ to $10^{-5}$ while checking convergence. The maximum dimension of the Hamiltonian matrix which we diagonalized during DMRG sweeps was of the order of $10^5$.

We also implemented single site corrections in the DMRG algorithm, which had been proposed by White in Ref.\cite{whitemart}, by admixing information from the Hamiltonian to the density matrix. This is necessary because the interactions in the projected Hamiltonian are not short-range.  These corrections improve the convergence by a considerable amount. Although in our calculations we used a version of the algorithm with a single central site, it is possible to extend it to two or more sites. However this would require more computer resources.

\section{Applications}

In this section we apply the method developed above to study systems of bosons with contact interactions
on the lowest Landau level in cases of the narrow channel geometry in the limit of large densities
and strong interactions as well as cylinder and a sphere. We discuss the convergence
of the method for different filling fractions in various geometries.

\subsection{Bosons in the narrow channel geometry in the strongly interacting regime}
The system of bosons in the narrow channel is described by Hamiltonian (\ref{ham_sec}) and was studied recently in Refs.\cite{matveenko,sinha} in the mean-field
regime in the limit of large density and weak interactions. It shows a series of phase transitions as a function of the
interaction strength or the particle density, between the states with different number of vortex rows. These states are gapless and have a Bogoliubov spectrum at small energies. In the opposite limit of very strong interactions bosons enter a state which is similar to the Laughlin $\nu=1/2$ state, in the case of contact interaction potential, which minimizes their interaction energy.
\begin{figure}[tb]
\epsfig{file=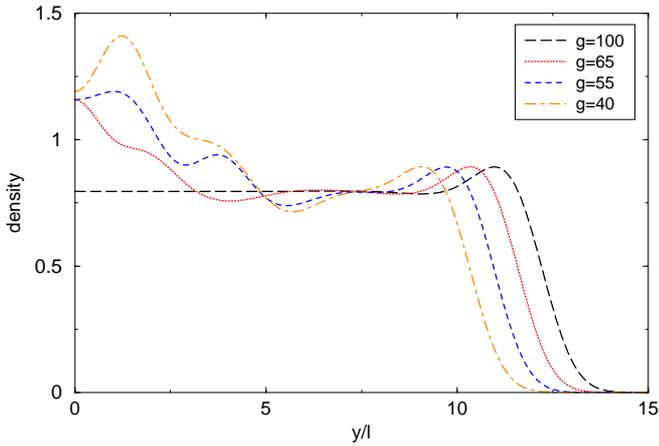,width=6cm,angle=270}
\caption{Density profiles for rotating bosons in the narrow channel geometry (in presence of a trap) for $N_p=20$, $L=10l$,  and different interaction strengths; long dashed line $g=100$, dotted line $g=65$ , short-dashed line $g=55$, dot-dashed line $g=40$.}
\label{fig:1}
\end{figure}
This state has a bulk gap and gapless edge states. When the interaction strength is decreased, the gap should close and there has to be be a phase transition at intermediate values of interactions. Notice that in the narrow channel the system area is increased with increasing interaction strength as bosons repel each other, which leads to the increase in the number of flux quanta penetrating the system and decrease in the
effective filling fraction $\nu$.\cite{chakraborty}

\begin{figure}[b]
\epsfig{file=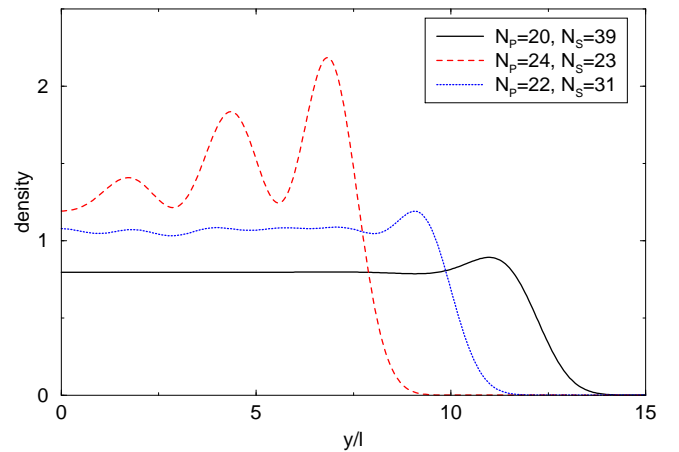,width=6cm,angle=270}
\caption{Density profiles for the cylinder geometry with $L=10l$ for different filling fractions; $\nu=1/2$ with $N_p=20$, $N_s=39$ solid line, $\nu=2/3$ with $N_p=22$, $N_s=31$ dotted line and $\nu=1$ with $N_p=24$, $N_s=23$ dot-dashed line.}
\label{fig:3}
\end{figure}

Here we calculate a few phases starting from very large interactions. The results for the integrated over $x$ particle density $n(y)$ , which is given by the equation
\begin{equation}
 n(y)=\frac{1}{\sqrt{\pi}l}\sum_{k}\bar{n}_k e^{-\frac{1}{l^2}(y-k l^2)^2},\label{ny}
\end{equation}
where $\bar{n}_k=\langle\hat{a}_k^{+}\hat{a}_k\rangle$, are presented in Fig.\ref{fig:1}. We take the system size $L=10l$, number of particles $N_p=20$ and up to $N_s=41$ momentum states in our calculations. For very strong interactions $g>68$ the ground state has a density profile which is very close to the one calculated for a cylinder geometry with $\nu=1/2$ and which is described by the Laughlin wave-function
\begin{equation}
 \psi(\{z_i\})=\left[\prod_{i<j} (e^{i\frac{2\pi}{L}z_i}-e^{i\frac{2\pi}{L}z_j})^2\right] e^{-\frac{1}{2}\sum_{i}y^{2}_{i}/l^2},
\end{equation}
see Fig.\ref{fig:3}. In this state the average occupation numbers $\bar{n}_k\sim1/2$ in the bulk and increase at the edges by a small amount. The total number of occupied k-states is equal to $N_s=2N_p-1$. When the interaction is decreased, the system enters another state through first order phase transition at $g\sim 68$. In this state the system shrinks by exactly two momentum states $\pm k_{max}$, where $k_{max}$ is the state with the highest momentum which is occupied in the quasi-Laughlin wave-function, and the extra particles move to the center of the trap which is seen as a bump in the density on the Fig. \ref{fig:1} at $g=65$. The area surrounding this bump continues to have the same constant density as before. This transition can be supported by the following argument. The energy cost of putting a particle to the center of the trap is of the order of $g$, and the gain of removing a particle from the state with momentum $k_{max}$ is of the order of $k_{max}^2$. The transition happens when $g\sim k_{max}^2$. Decreasing the interactions further we observe a series of phase transitions with the analogous behavior.\cite{full_diagram} This situation is similar to phase transitions in systems of lattice bosons in a trap, which shows wedding-cake structures which appear when changing the interaction strength or particle density. Similar structures have been studied by Cooper et. al.\cite{schoutens} in the case of quantum Hall effect.

\subsection{Cylinder geometry}\label{cylindersec}

Let us now consider a system of bosons on a cylinder. This geometry was discussed in the fermionic case by Rezayi, Haldane \cite{rezayi} and later studied by Bergholtz and Karlhede using DMRG in Ref.\cite{bergholtz}

We will impose periodic boundary conditions on the system of size $L$ along the $x$ axis, so that the momentum is quantized as $k=2\pi n/L$ where $n\in\mathbb Z$. The lowest Landau level basis wave-functions are given by the equation $(\ref{phik})$. This geometry could possibly be realized in experiments using proposals of creating artificial magnetic fields in optical lattices.\cite{osterloh} The number of momentum eigenstates is finite because of the finite extent $W$ of the bar in the $y$ direction and is given by $N_s=2N_{max}+1\sim L W/{2\pi l^2}$. The Hamiltonian of the system reads
\begin{equation}
 \hat{H}_c=\frac{1}{2}\sum_{ijkl}V_{ijk}\hat a_i^{+}\hat a_j^{+}\hat a_k \hat a_l\delta_{i+j,k+l},
\end{equation}
where matrix elements $V_{ijk}$ are given by the equation (\ref{matr_el}) with $g=g_{2d}/\sqrt{2\pi}Ll$ which plays a role of the energy scale.

\begin{figure}[tb]
\epsfig{file=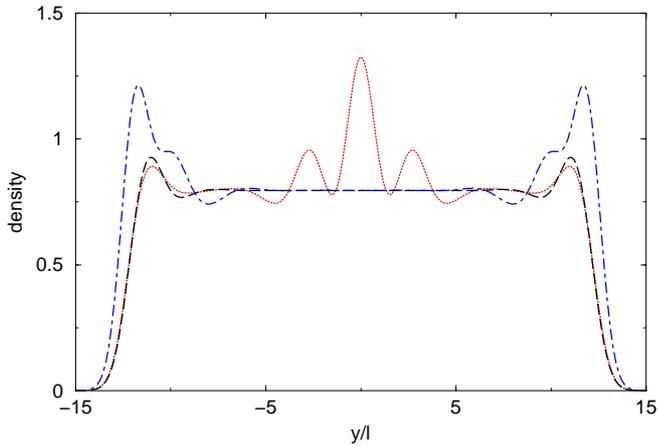,width=6cm,angle=270}
\caption{Density profiles for the cylinder geometry, $N_s=39$, $L=10l$ and $N_p=20$ (dashed line) which corresponds to $\nu=1/2$ Laughlin state, the same with one extra particle $N_p=21$ (dotted line) and with two extra particles $N_p=22$ (dot-dashed line).}
\label{fig:4}
\end{figure}
\begin{figure}[bt]
\epsfig{file=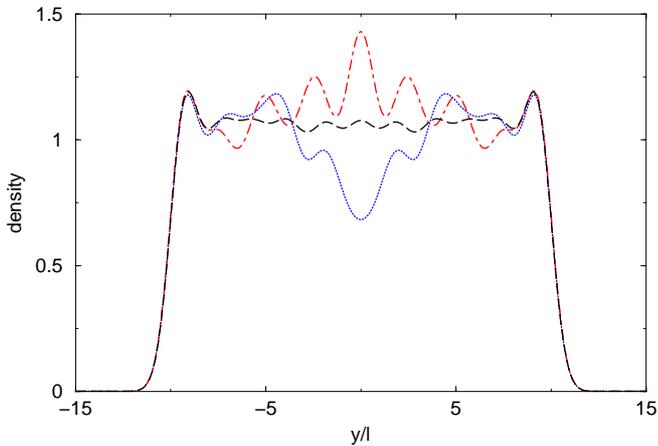,width=6cm,angle=270}
\caption{Density profiles for the cylinder geometry with $L=10l$ and $Ns=31$ for $N_p=22$ (dashed line) corresponding
to $\nu=2/3$ filling fraction, the same state with one particle removed $N_p=21$ (dotted line) and one particle
added (dot-dashed line).}
\label{fig:5}
\end{figure}

It is equivalent to the Hamiltonian of the narrow channel geometry with ``kinetic energy'' term set to zero.

\begin{figure}[t]
\epsfig{file=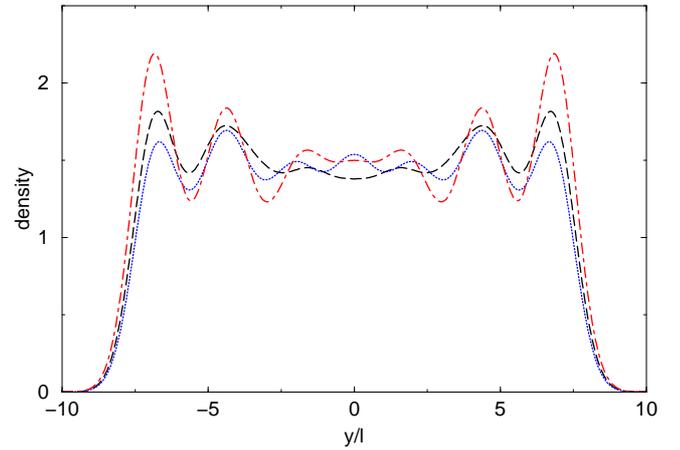,width=6cm,angle=270}
\caption{Density profiles for the cylinder geometry with $L=10l$ and $Ns=23$ for $N_p=24$ (dashed line) corresponding
to $\nu=1$ filling fraction, the same state with one particle removed $N_p=23$ (dotted line) and one particle
added $N_p=25$ (dot-dashed line).}
\label{fig:7}
\end{figure}

We calculated the density profiles (\ref{ny}) for systems of length $L=10l$ and different filling fractions, which is
shown in Fig. \ref{fig:3}. The energy of the $\nu=1/2$ Laughlin state is zero, the best energies per particle of the $\nu=2/3$ state on a cylinder $E_{\nu=2/3}\approx 0.489853g$ and for the $\nu=1$ state $E_{\nu=1}\approx 1.199865g$. These energies have been calculated with relative precision $\approx 10^{-5}$, however convergence of the problem is nonlinear and the error is difficult to estimate. We have also calculated the energies and the density profiles for the states with extra added or removed particles which is presented in Figs. \ref{fig:4},\ref{fig:5},\ref{fig:7}. For the Laughlin $\nu=1/2$ state the energy of the state with one extra particle is equal to $\Delta E_{\nu=1/2}^{(+1)}=2.1917252g$ which corresponds to $N_p=21$ and for two particles is given by $\Delta E_{\nu=1/2}^{(+2)}=3.5763744g$ for $N_s=39$ and $L=10l$. For the $\nu=2/3$ state the energy cost of adding one particle is $\Delta E_{\nu=2/3}^{(+1)}=2.532379g$ and the energy change for removing a particle $\Delta E_{\nu=2/3}^{(-1)}=-1.5541128g$, here $L=10l$ and $N_s=31$. The energy change for the $\nu=1$ state with added and removed particle are given by $\Delta E_{\nu=1}^{(+1)}=3.66319g$ and $\Delta E_{\nu=1}^{(-1)}=-3.29027g$ with $N_p=25$ and $N_p=23$ particles correspondingly for $N_s=23$ and $L=10l$.

In general the properties of the system will depend on the length $L$ and to approach a thermodynamic limit one have to send $L$ to infinity. In the case of $\nu=1/2$ finite-size effects in the density profile are already very small for $L=10$, although one can still see oscillations for the states with higher filling fractions $\nu=2/3$ and $\nu=1$. This could be an effect of the finite system length $L$ or finite number of particles. In order to separate these two contributions it would be necessary to perform calculations for systems of larger size. It is important to mention that with increasing $L$ the interactions between particles in different momentum states will increase which will result in decrease of the numerical convergence and will require more computer resources. However, it can be shown in certain cases that important physics can be inferred from studies of systems with finite lengths. The authors of paper Ref.\cite{rezayi} observed that a system of fermions on a cylinder undergows continuous transformation, i.e. without a phase transition, from the Laughlin to the Tao-Thoules state (see also Ref.\cite{hansson}). Similar results were obtained in Ref.\cite{bergholtz} for $\nu=1/2$ state of electrons on a cylinder and in more general case on a thin torus in Ref. \cite{bergholtztr} One of the interesting applications of our method would be to study this adiabaticity in the bosonic case.

\subsection{Spherical geometry}
Let us now turn to the case of the spherical geometry. Exact diagonalization on a sphere, which was introduced by Haldane in Ref.\cite{Hal} is one of the most used in the finite-size studies of fractional quantum Hall effect for both bosons and fermions. Application of DMRG to systems of fermions on a sphere was developed in Ref.\cite{feiguin_recent}, here we consider bosonic case.

The Hamiltonian of the system of bosons with delta-function interaction in the spherical geometry is given by the equation
\begin{equation}
 \hat{H}_s=\frac{1}{2}\sum_{m_{i}}V_{\{m_{i}\}}\hat{a}^{+}_{m_1}\hat{a}^{+}_{m_2}\hat{a}_{m_3}\hat{a}_{m_4}\delta_{m_1+m_2,m_3+m_4},
\end{equation}
where the matrix elements of interactions on the lowest Landau level read
\begin{equation}
 V_{\{m_i\}}=2\frac{g}{4\pi S}\frac{(2S+1)^2}{4S+1}
\frac{[\Pi_{i=1}^{4}C^{2S}_{S+m_i}]^{1/2}}{C^{4S}_{2S+m_1+m_2}}.
\end{equation}
Here $C^{n}_{k}=n!/k!(n-k)!$ are binomial coefficients and $S\equiv N_{max}$ which gives $N_s=2S+1$.

\begin{figure}[tbh]
\epsfig{file=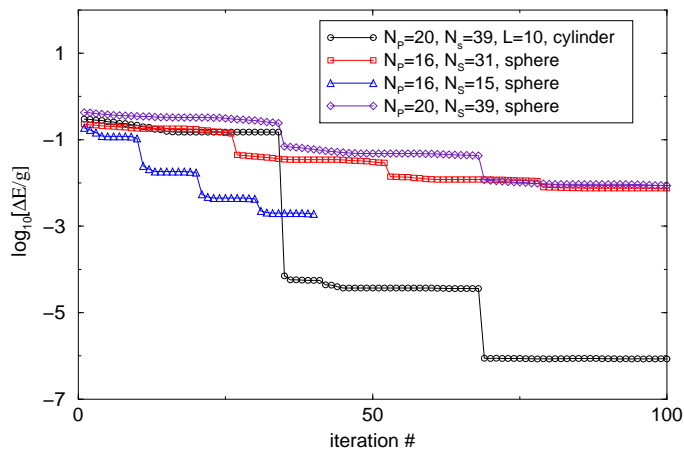,width=6cm,angle=270}
\caption{Absolute error in the ground state energy for the Laughlin state $\nu=1/2$ on a cylinder with $L=10l$, $N_p=20$,
$N_s=39$ (circles), $\nu=1/2$ Laughlin state on a sphere for $N_p=16$, $N_s=31$ (squares), $N_p=20$, $N_s=39$ (diamonds), $\nu=1$ on a sphere for $N_p=16$, $N_s=15$ (triangles) as a function of the number of iterations.}
\label{fig:conv}
\end{figure}

In this paper we studied convergence of the ground state energies at filling fractions $\nu=1$ and $\nu=1/2$ on a sphere. The results are presented in Fig. \ref{fig:conv} together with the convergence for the $\nu=1/2$ Laughlin state in the cylinder geometry for comparison. In the calculations we used the states of the density matrix with the eigenvalues larger than $\varepsilon=10^{-8}$. We find that for a cylinder with $L=10l$, convergence is generally much better than that of the spherical geometry. This is because the matrix elements of the Hamiltonian on a cylinder for this value of $L$ fall-off much faster with the distance between sites in the momentum space than for a sphere (see also discussion in Sec.\ref{cylindersec}).

\section{Conclusions}
In conclusion, we developed a momentum space density matrix renormalization group (DMRG) technique to study bosonic fractional quantum Hall effect. Using this method in the narrow channel geometry we found quantum phase transitions out of the $\nu=1/2$ Laughlin-type state at the critical value of the interaction strength. We calculated energies of ground states as well as states with added or removed particles and the density profiles at filling fractions $\nu=1/2$, $\nu=2/3$ and $\nu=1$ on a cylinder. The number of particles in our calculations is considerably larger than in current exact diagonalization studies of bosonic QHE. We found that convergence of the method strongly depend on geometry and it requires more computational efforts to study systems on a sphere, compared to a cylinder, provided that cylinder is thin enough (which is always the case in our calculations).

\acknowledgments{I am grateful to Gora Shlyapnikov, N.~R.~Cooper, Th.~Jolicoeur, N.~Regnault and S.~Ouvry for numerous fruitful discussions. My work was supported by the \mbox{EPSRC} grant EP/D066379/1. I would like to thank Laboratoire de Physique Th\'{e}orique et Mod\`{e}les Statistiques in Orsay for the hospitality, where a part of this work was done. I acknowledge MPIPKS, Dresden for computer time which I used during my visits.}


\begin{thebibliography}{0}
\bibitem{review} U.~Schollw\"{o}ck, Rev. Mod. Phys. \textbf{77}, 259 (2005).
\bibitem{dynamics} S.~R.~Manmana, S.~Wessel, R.~M.~Noack, and A.~Muramatsu, Phys. Rev. B \textbf{79}, 155104 (2009).
\bibitem{fendley} A.~Feiguin, P.~Fendley, M.~P.~A.~Fisher, and C.~Nayak, Phys. Rev. Lett. \textbf{101}, 236801 (2008).
\bibitem{tensor} Z.-C.~Gu, M.~Levin, B.~Swingle, and X.-G.~Wen, Phys. Rev. B 79, 085118 (2009);
Z.-C.~Gu, M.~Levin, and X.-G.~Wen, Phys. Rev. B 78, 205116 (2008);
Z.~Y. Weng, and T.~Xiang, Phys. Rev. Lett. \textbf{101}, 090603 (2008).
\bibitem{vidal} L.~Tagliacozzo, G.~Evenbly, G.~Vidal, arXiv:0903.5017.
\bibitem{shibata} N.~Shibata and D.~Yoshioka, Phys. Rev. Lett. \textbf{86}, 5755 (2001);
N.~Shibata and D.~Yoshioka , J. Phys. Soc. Jpn. \textbf{72}, 664 (2003).
\bibitem{feiguin_recent} A.~E.~Feiguin, E.~Rezayi, C.~Nayak, S.~Das Sarma, Phys. Rev. Lett. \textbf{100}, 166803 (2008).
\bibitem{feiguin}A.~E.~Feiguin, E.~Rezayi, Kun Yang, C.~Nayak, S.~Das Sarma, Phys. Rev. B. \textbf{79}, 115322 (2009).
\bibitem{papenbrock}J. Dukelsky and S. Pittel, Rept. Prog. Phys. {\bf 67} 513 (2004); T. Papenbrock and D.J. Dean, J. Phys. G: Nucl. Part. Phys. {\bf 31} S1377 (2005).
\bibitem{ketterle}C.~H.~Schunck, M.~W.~Zwierlein, A.~Schirotzek, and W.~Ketterle, Phys. Rev. Lett. \textbf{98}, 050404 (2007); J.~R.~Abo-Shaeer,  C.~Raman,  J.~M.~Vogels,  W.~Ketterle, Science, Vol. \textbf{292}. no. 5516, pp. 476 
\bibitem{cooper} N.~R.~Cooper, N.~K.~Wilkin, and J.~M.~F.~Gunn, Phys. Rev. Lett. \textbf{87}, 120405 (2001).
\bibitem{cornell} V.~Schweikhard, I.~Coddington, P.~Engels, V.~P.~Mogendorff, and E.~A.~Cornell, Phys. Rev. Lett. 
\textbf{92}, 040404 (2004).
\bibitem{ho} Tin-Lun Ho, Phys. Rev. Lett. {\bf 87}, 060403 (2001).
\bibitem{osterloh} K. Osterloh, M. Baig, L. Santos, P. Zoller, and M. Lewenstein, Phys. Rev. Lett. \textbf{95}, 010403 (2005); J. Ruseckas, G. Juzeliunas, P. \"Ohberg, and M. Fleischhauer, Phys. Rev. Lett. \textbf{95}, 010404 (2005).
\bibitem{lin} Y.-J.~Lin, R.~L.~Compton, A.~R. Perry, W.~D.~Phillips, J.~V.~Porto,
and I. B. Spielman, Phys. Rev. Lett. \textbf{102}, 130401 (2009).
\bibitem{matveenko} S.~I.~Matveenko, D.~Kovrizhin, S.~Ouvry, G.~V.~Shlyapnikov, Phys. Rev. A \textbf{80}, 063621 (2009).
\bibitem{regnault} N.~Regnault and Th.~Jolicoeur, Phys. Rev. B \textbf{69}, 235309 (2004); N. Regnault et. al., J. Phys. B: At. Mol. Opt. Phys. 39 (2006) S89–S99; C.-C.~Chang, N.~Regnault, Th.~Jolicoeur, and J.~K.~Jain, Phys. Rev. A \textbf{72}, 013611 (2005).
\bibitem{ladders}S.~R.~White, D.~J.~Scalapino, Phys. Rev. Lett. \textbf{91}, 136403 (2003).
\bibitem{gora1} D.~S.~Petrov, M.~Holzmann, and G.~V.~Shlyapnikov, Phys. Rev. Lett. {\bf 84}, 2551 (2000).
\bibitem{white} S.~R.~White, Phys. Rev. Lett. {\bf 69} 2863 (1992); S.~R. White, Phys. Rev. B \textbf{48}, 10345 (1993).
\bibitem{xiang} T.~Xiang, Phys. Rev. B {\bf 53}, R10445 (1996).
\bibitem{schoutens} N.~R.~Cooper, F.~J.~M.~van~Lankvelt, J.~W.~Reijnders and K.~Schoutens,
Phys. Rev. A {\bf 72}, 063622 (2005).
\bibitem{bergholtz} E.~J.~Bergholtz and A.~Karlhede, arXiv:cond-mat/0304517
\bibitem{hansson} T.~H.~Hansson and A.~Karlhede, arXiv:0907.0672
\bibitem{bergholtztr} E.~J.~Bergholtz and A.~Karlhede, Phys. Rev. Lett. \textbf{94}, 026802 (2005); E.~J.~Bergholtz and A.~Karlhede, Phys. Rev. B \textbf{77}, 155308 (2008).
\bibitem{full_diagram} the full phase diagram will be discussed elsewhere.
\bibitem{whitemart} S.~R.~White, L.~R.~Martin,  The Journal of Chemical Physics, {\bf 110}, 4127 (1999).
\bibitem{sinha} S.~Sinha ahd G.~V.~Shlyapnikov, Phys. Rev. Lett. {\bf 94}, 150401 (2005).
\bibitem{chakraborty} T.~Chakraborty, K.~Niemela and P.~Pietilainen, Phys. Rev. Lett. {\bf 78}, 4829 (1997).
\bibitem{rezayi} E.~H.~Rezayi, F.~D.~M.~Haldane, Phys. Rev. B {\bf 50}, 17199 (1994).
\bibitem{Hal} F.~D.~M.~Haldane, Phys. Rev. Lett. \textbf{51}, 605 (1983).
\bibitem{whiterev} S.~R.~White, Phys. Rev. B {\bf 72}, 180403(R) (2005).
\bibitem{neder} I.~Neder, M.~Heiblum, Y.~Levinson, D.~Mahalu, and V.~Umansky, Phys. Rev. Lett. \textbf{96}, 016804 (2006);
D.~L.~Kovrizhin and J.~T.~Chalker, Phys. Rev. B {\bf 80}, 161306(R) (2009).
\end{thebibliography}
\end{document}